# EVANESCENT OSCILLATIONS IN ACCELERATING STRUCTURES

**Ayzatsky M.I.[1]**
NSC KIPT, Akademicheskay str., 1, Kharkov, Ukraine, 61108
aizatsky@nik.kharkov.ua

It is well known, that in infinite periodic structures there are two different electromagnetic eigen-oscillations supported by medium without external currents and charges. In the certain frequency intervals (passbands) the electromagnetic oscillations represent a wave process, which carry constant energy in direct or opposite directions. Between the passbands the electromagnetic oscillations have a structure that is distinct from previous case. In these frequency intervals electromagnetic oscillations transfer no energy in the direction of periodicity and have decreasing (increasing) dependence on the coordinate and they are called evanescent waves (oscillations). These frequency intervals are called forbidden bands (stopbands). Today the 1-D periodic mediums are sometimes called 1-D photonic band gap (PBG) structures. Evanescent oscillations that are exited in finite structures can be treated in the terms of the so-called defect modes as the "surface defect levels" that arise as the result of existence of the interface between the medium with a band structure and the mediums that can be considered as an infinite potential wall. Results of our investigations of the properties of the electromagnetic oscillations in the resonators based on the periodic waveguides are represented. These results point out on the possibility of existing in resonant cavities the new type of eigen oscillations based on the evanescent waves. Such oscillations do not exist in the smooth waveguides. They have a longitudinal distribution of electromagnetic field with decreasing (increasing) dependence on the coordinate. We discuss the using of such oscillations for bunching and accelerating of electron beams.

### Introduction

In the case of a smooth waveguide at a fixed frequency, number of variations along a radius and an azimuth angle there are two electromagnetic eigen-oscillations that describe either two waves propagating in opposite directions or evanescent oscillations. If we put bounds in the form of two parallel metal screens in such waveguide we shall obtain a resonator. Its properties will be depend on the distance between the sheets only and can be explained on the base of the interference process of the two propagating waves (see, for example, [1,2]). This circumstance is defined by a fact that the plane wavefront of the eigen propagating waves gives the possibility to fulfill the boundary conditions with the two propagating waves without taking into account evanescent oscillations and there are no conditions when the evanescent oscillations could be exited. From this follows that the eigen modes of resonator can exist above the cutoff frequencies of the infinite waveguide and the boundary conditions on the lateral surfaces give the possible values of the wavelength $\Lambda$: $\Lambda = L \times 2/n$, where $L$ - the length of the resonator. Knowing the dispersion characteristic of a waveguide $\omega = f_i(h)$ ($h$- the wavenumber, $i$ –the set of numbers that characterize the transversal distribution), we can find the eigen frequencies: $\omega_{k,i} = f_i(2\pi k/L)$.

In the case of forming a resonator on the base of a symmetric periodic waveguide[2], the pointed above reasoning become more complicated. The eigen waves of a symmetric periodic waveguide exist in the definite pass bands and have a plane wave front only in the symmetry planes. As a result, the eigen oscillations of the resonator which is based on a periodic waveguide can be described as a result of summing two traveling waves under the condition when the first metal screen is placed in the symmetry plane and the second – at the distance which is an integer number of the waveguide period $L = N \times D$. Under these conditions the eigen frequencies of such resonators, which we shall call "symmetric", lay on the dispersion curves of propagating waves with the values of longitudinal wavenumber that is defined by the number of periods $N$. This circumstance is widely used in the procedure of cell tuning under the manufacturing the periodic waveguides (see, for example, [3]). Under the arbitrary placing of the first metal sheet relatively the symmetry plane (such resonators we shall call "nonsymmetric") we cannot fulfill the boundary conditions with the two propagating waves without taking into account evanescent oscillations. So, the eigen frequencies of the nonsymmetric resonators will be differ from the eigen frequencies of the symmetric ones. Is there a possibility for the eigen frequencies to lay outside the pass band of an infinite periodic waveguide? In this case such eigen modes can not be defined as superposition of the two propagating waves and will be defined as superposition of a set of evanescent oscillations. In this case electromagnetic fields will be have increasing (decreasing) along the longitudinal coordinate spatial distributions.

---

[1] Aizatsky N.I.
[2] Under symmetric periodic waveguides we shall mean waveguides in which one period has the symmetry plane along the coordinate z, in this case a waveguide will have two symmetry planes.

### 1. Eigen frequencies of a resonator based on a symmetric periodic waveguide

The number of the eigen frequencies of the symmetrical resonator within the definite pass band depends on the plane of symmetry in which the metal sheets are placed. Indeed, the symmetric periodic waveguides have two planes of symmetry: the first one is a cross-section of a waveguide in the middle of the cavity and the second one - a cross-section of a waveguide in the middle of coupling aperture. When the metal sheets are placed in the middle of the cavity and the distance between the sheets equals $L = N \times D$, the number of eigen frequencies will be equal ($N+1$) and the eigen frequencies we can find from the equation $h_i(\omega) \times D = \pi k / N$, where the number $k$ can take the values $k = 0, 1, ..., N$. Such resonators we shall call symmetric resonators of the first kind. When the metal sheets are placed in the middle of the coupling aperture the number of eigen frequencies will be equal $N$ and the eigen frequencies we can find from the above equation, where the number k can take the values $k = 0, 1, ..., N-1$. In such resonators we cannot excite the "π"-type mode. Such resonators we shall call symmetric resonators of the second kind. But the symmetric resonator of the second kind can be obtain from the symmetric resonator of the first kind by gradual moving of the metal sheets relatively the plane of symmetry of the periodic waveguide and keeping constant the distance between the sheets $L = N \times D$. So, there can arise such question: in what type does the oscillation of the "π"-type of symmetric resonators of the first kind transform under such moving and what states will be intermediate.

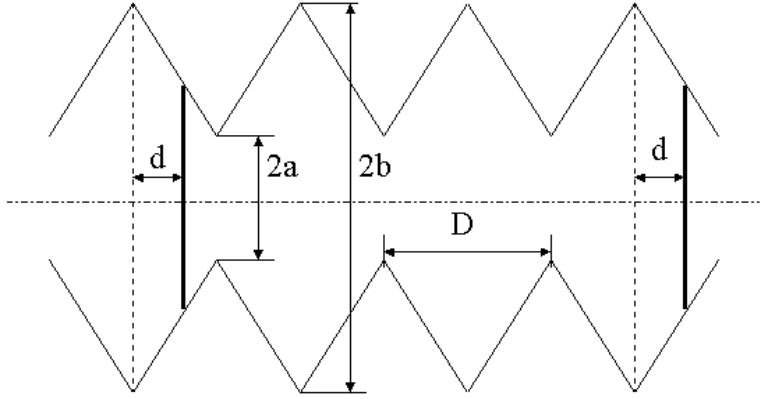

Figure 1

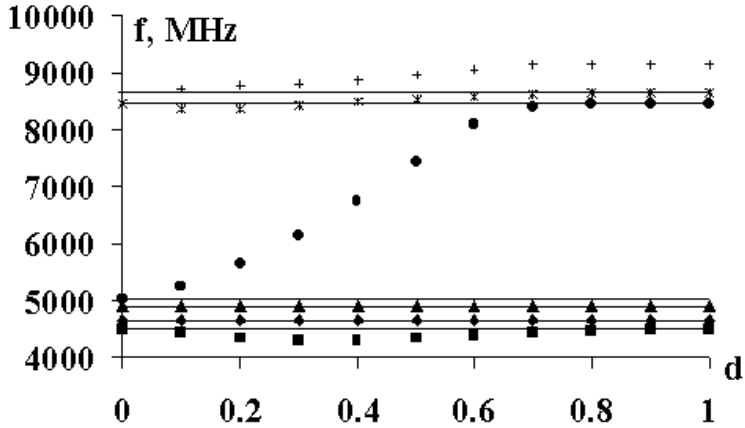

Figure 2

Using the SUPERFISH code we made the simulation of the dependencies of the eigen frequencies of the resonator (which have been formed on the base of a symmetric periodic waveguide depicted on Fig.1) on the distance *d* between the first plane of symmetry and the first metal sheet under keeping constant the distance between the sheets $L = 3 \times D$ (*D*=2 cm, *a*=1.25 cm, *b*=3.5 cm). From Fig.2, which shows these dependencies, we can see that the "π"-type oscillation of the symmetric resonator of the first kind which belong to the $E_{01}$-passband of the waveguide under smooth moving of the metal sheets relatively the plane of symmetry transforms in the "0"-type oscillation of the symmetric resonator of the second kind which belong to the $E_{02}$-passband of the periodic waveguide with two radial variations. All eigen frequencies of the symmetric resonator of the first kind which belong to the $E_{02}$-passband of the periodic waveguide under smooth moving of the metal sheets shift up. This means that these eigen oscillations change the transversal type $E_{02k} \to E_{02k+1}$. Eigen oscillations of the symmetric resonator of the first kind, which belong to the $E_{01}$-passband, do not transfer their type.

So, the results of our simulation show that under smooth transforming of the symmetric resonator of the first kind into the symmetric resonator of the second kind through the set of the nonsymmetric resonators, some eigen frequencies ("π"and "0"-types) of the nonsymmetric resonators lay in forbidden zones: "π"-type -- in the first forbidden zone and "0"-type – in the main forbidden zone. These results point out on the possibility of existing in resonant cavities the new type of eigen oscillations based on the evanescent waves. Such oscillations do not exist in the smooth waveguides. They have a longitudinal distribution of electromagnetic field with decreasing (increasing) dependence on the coordinate.

In the considered above case eigen oscillations with the frequency greater than the frequency of "π"-type wave of an infinite periodic waveguide are formed on the base of an infinite set of evanescent waves. Apparently, such eigen oscillation can be formed on the base of one evanescent oscillation, as in the case of a layered dielectric [4,7]. For realization such distribution it is necessary to find a special form of the first and last cells, which give the possibility to fulfill the boundary conditions along a complicated cross-section for supporting only one evanescent oscillation.

### 2. Using new type oscillations in accelerators

As we have already said above, the existence of the eigen oscillations based on evanescent waves of an infinite periodic waveguide gives the possibility to create increasing dependence of the electric field on the coordinate. Such distribution of the electric field allows to improve the capture and bunching process at the initial stage of accelerating. We have recently shown [4] that the eigen oscillations based on one evanescent wave can be created in the bi-periodic waveguide which has the additional small forbidden gap between two pass bands. Such systems were used in some injector structures [5,6]. As the excitation of such cavity stack is performed through the last resonator the particles of electron beam are bunched and accelerated in the growing field. In the case of a bi-periodic waveguide the main evanescent wave have the structure that is similar, except growing, to the one that is formed by two traveling waves with the phase shift per sell equals π/2 and a zero value amplitude in even cells. Existence of cells with amplitudes equal to zero amplify the longitudinal bunching which became similar to the same process in klystrons.

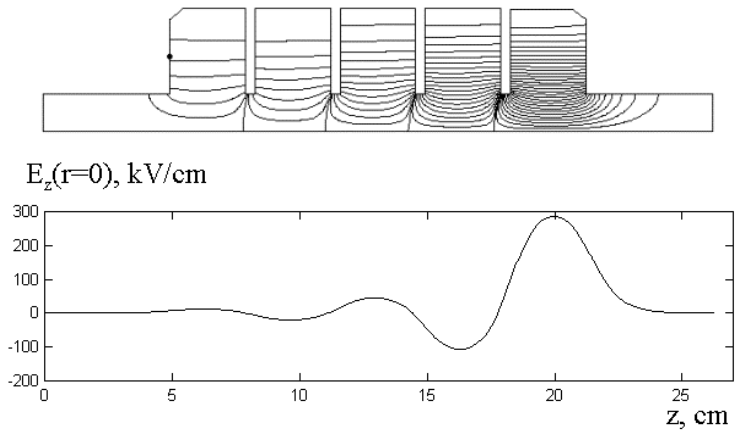

Figure 3

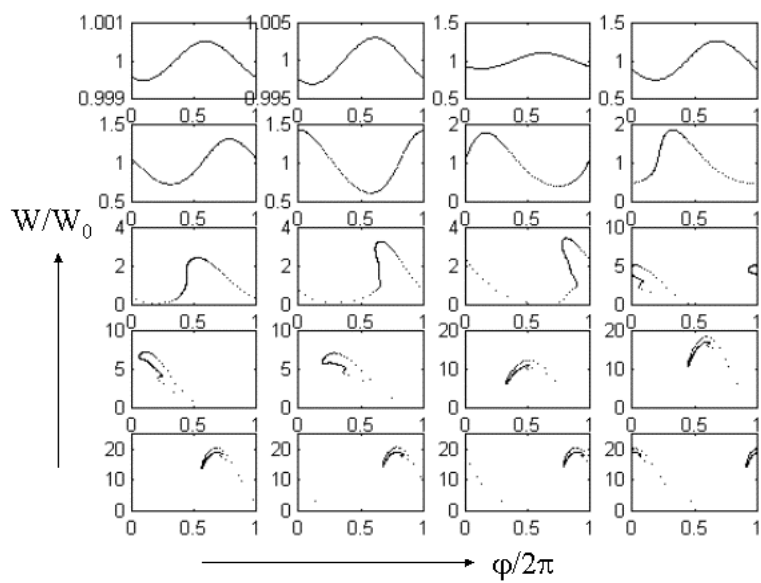

Figure 4

The above results show that we can use the resonant system which is based on the segment of a periodic waveguide terminated with special end cells and which eigen frequencies are greater (under electric coupling) than the cutoff frequency of "π"-type of an infinite waveguide. In this case the phase shift between the cells will be equal π. As we have already said, for creating a resonator which field will be based on one evanescent wave of an infinite waveguide we must know the structure of this evanescent wave and choose such form of end cells which will support only this evanescent wave. This task has not been solved up to now. But we can use the eigen oscillation which is based on the set of evanescent waves, as in the considered above case of the plane transverse boundaries. The simplest way to create such growing distribution is such: at the first stage we create a resonator chain which has "π"-type eigen oscillation with full end cells (see, for example [8]) and then shift up the frequency of the last end cell (see Fig.3).

Simulation of the particle dynamic in such structure shows that the bunching and accelerating processes are effective and we can obtain bunches with a small longitudinal size (see Fig.4, where the bunching process of a beam without preliminary modulation is shown on the energy-phase plane for different values of the transversal coordinate $z_s=L/20 \times s$; the initial beam energy equals $W_0$=50 keV).

### 3. Acknowledgement
The author wish to thank Kushnir V.A. and Mitrochenko V.V. for helpful discussion.